\documentstyle[prl,aps,multicol]{revtex}
\renewcommand{\narrowtext}{\begin{multicols}{2} \global\columnwidth20.5pc}
\renewcommand{\widetext}{\end{multicols} \global\columnwidth42.5pc}
\def\top#1{\vskip #1\begin{picture}(290,80)(80,500)\thinlines \put(
65,500){\line( 1, 0){255}}\put(320,500){\line( 0, 1){
5}}\end{picture}}
\def\bottom#1{\vskip #1\begin{picture}(290,80)(80,500)\thinlines \put(
330,500){\line( 1, 0){255}}\put(330,500){\line( 0, -1){
5}}\end{picture}}

\begin{document}
\draft

\title{Theory of Exciton Recombination from the Magnetically Induced
Wigner Crystal}

\author{N. R. Cooper\cite{newaddress}}

\address{Lyman Laboratory of Physics, Harvard University, Cambridge,
MA 02138, USA.}

\date{8 December, 1995}

\maketitle

\begin{abstract}
We study the theory of itinerant-hole photoluminescence of
two-dimensional electron systems in the regime of the magnetically
induced Wigner crystal.  We show that the exciton recombination
transition develops structure related to the presence of the Wigner
crystal.  The form of this structure depends strongly on the
separation $d$ between the photo-excited hole and the plane of the
two-dimensional electron gas.  When $d$ is small compared to the
magnetic length, additional peaks appear in the spectrum due to the
recombination of exciton states with wavevectors equal to the
reciprocal lattice vectors of the crystal.  For $d$ larger than the
magnetic length, the exciton becomes strongly confined to an
interstitial site of the lattice, and the structure in the spectrum
reflects the short-range correlations of the Wigner crystal.  We
derive expressions for the energies and the radiative lifetimes of the
states contributing to photoluminescence, and discuss how the results
of our analysis compare with experimental observations.

\end{abstract}



\narrowtext

\section{Introduction}

Recent experimental studies of high-mobility GaAs/AlGaAs devices using
itinerant-hole photoluminescence have shown this to be a sensitive probe of the
two-dimensional electron systems formed in these devices\cite{clarkreview}.
Features in the photoluminescence spectra have been found to correlate well with
transport measurements of the integer and fractional quantum Hall
effects\cite{turberprl,dahlheimansuppression,goldbergoverview}, and the
insulating regime observed at very low filling
fraction\cite{goldys,turberfieldeuro,heiman}.  These studies are particularly
interesting in this low filling fraction regime, where transport measurements
can provide only limited information on the state of the two-dimensional
electron gas.  In particular, they may provide information on the transition to
the Wigner crystal which is the expected groundstate under these conditions.
Related experimental studies of the photoluminescence of intentionally
acceptor-doped samples, in which the photo-excited hole becomes bound to an
acceptor, have also found structure associated with the transition to an
insulating regime\cite{buhmanncrystal,kuktriangle}.  However, the signature of
the insulating state in those experiments is quite different from that in the
undoped devices in which the photo-excited hole is free to move
(``itinerant-hole'' photoluminescence).

In the undoped samples, the transition to the insulating regime is
associated with the appearance of additional lines in the
photoluminescence spectrum\cite{goldys,turberfieldeuro,heiman}. It has
been suggested that the appearance of these lines does indeed indicate
the formation of a magnetically induced Wigner
crystal\cite{goldys,turberfieldeuro}.  However, the theory underlying
itinerant-hole photoluminescence in the extreme quantum limit is not
well understood.  Existing theories of photoluminescence in the Wigner
crystal regime apply only to the limit in which the photo-excited hole
is far from the electron gas compared to the typical electron-electron
spacing\cite{fertigpr,fertigshakeup}.  This condition is not
appropriate for typical devices in which itinerant-hole
photoluminescence is observed: while the asymmetry of these devices
does cause the hole to lie some distance from the electron gas, it is
believed to be close compared to the electron-electron
spacing\cite{vietbirman}.

We present a theory for itinerant-hole photoluminescence in the Wigner
crystal regime which applies to systems in which the photo-excited
hole is close to the two-dimensional electron gas.  We study the limit
of strong magnetic field in which all the electrons and the
photo-excited hole are confined to states in the lowest Landau
level. To represent the asymmetry of the quantum well, the hole is
assumed to lie in a plane a distance $d$ away from the two-dimensional
electron gas (which is also assumed to have zero thickness).  Such a
model is common in theories of photoluminescence in the fractional
quantum Hall regime\cite{apalkovprb,chenprb94,zangbirman}. In that
regime, the model presents an intrinsically strongly-coupled many-body
problem, which, for the most part, has required numerical
investigation.  For very low filling fraction and for $d$ small
compared to the typical electron-electron spacing, certain
simplifications arise.  Under these conditions, we expect that the
low-lying energy states in the presence of the photo-excited hole
involve the formation of an exciton, with one electron strongly bound
to the photo-excited hole.  A well-defined exciton can form provided
the size of the exciton, set by the magnetic length
$l\equiv\sqrt{\hbar/eB}$, is small compared to the electron-electron
spacing.  Moreover, since the exciton is neutral, it will couple
rather weakly with the other electrons.  In particular, we expect that
for small $d$, the presence of the exciton will not significantly
perturb the groundstate (Wigner crystal) configuration of the other
electrons, and the exciton will behave as a rather non-invasive probe
of this crystalline state.  This expectation is motivated by studies
of the classical groundstate of a system of electrons in the presence
of an ionized donor impurity located a distance $d$ from the plane of
the two-dimensional electron gas\cite{classicalruzin}. It is found
that for $d<0.29 a$ (where $a$ is the lattice constant of the Wigner
crystal) the system adopts a groundstate configuration in which one
electron is bound to the donor impurity, while the remaining electrons
lie close to the sites of a triangular lattice.

In this paper, we discuss how the photoluminescence spectrum arising
from the recombination of the exciton is affected by the presence of
the crystal.  We study a model in which the exciton moves in the
static potential set up by a triangular lattice of electrons. Our
analysis neglects the dynamical properties of the lattice (phonon
coupling). This approximation is justified, as the structure that we
will find is on a energy scale large compared to the typical
magnetophonon energy, ${\cal O}(l^2 e^2 /\epsilon a^3)$: the lattice
can therefore be considered to be static on the timescale necessary to
define these exciton states.  In the present work we neglect exchange
processes between the electron in the exciton and those forming the
lattice; these will be considered in a separate
paper\cite{cooperzhang}.

The behaviour of our model is strongly dependent on the separation
$d$.  For small $d$, the coupling of the exciton to the lattice is
weak; in this case, additional peaks appear to higher energy of the
main exciton line arising from the recombination of exciton states
with momenta equal to reciprocal lattice vectors of the crystal. The
spectrum effectively carries diffraction information on the crystal
and is therefore sensitive to long-range crystalline order.  As $d$
increases, the coupling becomes stronger and the low-lying exciton
states become increasingly confined to the interstitial regions of the
crystal. These states are less sensitive to the presence of
crystalline order, but depend on the short-range correlations of the
electron gas.  The crossover from weak to strong coupling occurs when
the (dipole) potential energy of the exciton, ${\cal O}(d^2
e^2/\epsilon a^3)$, becomes larger than the kinetic energy cost to
confine the exciton to a size of order the lattice constant $\hbar^2
/M a^2$ (where $M$ is the effective mass of the exciton).  We show
that this crossover occurs when the separation between electron and
hole planes $d$ becomes larger than the magnetic length.

We compare the results of our model with experimental observations of
photoluminescence in the extreme quantum limit.  We find that the
energy scale predicted by our model for the splitting of the exciton
transition is comparable to the energy scale of the structure observed
in experiment.  We argue that the observed peak-splitting that has
been associated with the Wigner crystal regime could arise from the
groundstate and first excited state of a strongly-confined
interstitial exciton.  However, other explanations for this splitting
cannot be ruled out, and it may be that the structure that we predict
is not seen due to a lack of population of the higher energy exciton
states. Further experimental work is required in order to identify
this structure. We argue that this could best be observed in optical
absorption experiments.

The paper is organized as follows.  In the following section we review
the theory of the free two-dimensional exciton in a strong magnetic
field. We derive expressions for the binding energy and effective mass
of the exciton for the case in which the electron and hole lie in
planes separated by $d$.  In section~\ref{sec:weakcoupling}, we study
the coupling of the exciton to the periodic potential set up by a
Wigner crystal of other electrons.  We derive exact expressions for
the matrix elements of the periodic potential of the crystalline
lattice within the basis of free exciton states, and study the
resulting low-energy exciton states in the weak-coupling limit, $d\ll
l$, using perturbation theory about the free exciton states.  In
section~\ref{sec:interstitial} we study the limit of strong coupling,
$d\gtrsim l$.  To do so, we develop an effective Hamiltonian for the
motion of the exciton in a smooth external potential. We apply this to
the case in which the potential is due to the presence of a Wigner
crystal, and study the low-energy eigenstates close to an interstitial
site of the electron crystal.  For both the weak and strong coupling
limits we derive expressions for the energies and relative radiative
lifetimes of the low-lying energy states.  In
section~\ref{sec:comparison} we discuss how these results compare with
existing photoluminescence measurements on GaAs/AlGaAs
heterostructures in the extreme quantum limit.  Finally,
section~\ref{sec:conclusions} contains a summary of the main points of
the paper.

\section{Free Two-Dimensional Exciton States in Strong Magnetic Field}
\label{sec:freeexciton}

The wavefunctions and the energy spectrum of a free two-dimensional
exciton in a strong magnetic field have been discussed by Lerner and
Lozovik\cite{lernloz}.  They built on the work of Gor'kov and
Dzyaloshinskii\cite{gorkov} who showed that, within the effective mass
approximation for the conduction and valence bands, the exciton states
can be described in terms of a conserved two-dimensional momentum,
$\bbox{P}$.  In the limit of strong magnetic field, when the cyclotron
energies of both the electron and hole become large compared to the
typical electron-hole interaction energy, the exciton wavefunction is
completely specified by the momentum $\bbox{P}$ and the Landau level
indices of the electron and hole.  The lowest energy exciton state, in
which the electron and hole are both in the lowest Landau level, has
the wavefunction\cite{gorkov} \widetext
\top{-2.8cm}
\begin{equation}
\label{eq:llwavefunctions} 
\langle \bbox{r}_e, \bbox{r}_h|\bbox{P}\rangle = \frac{1}{\sqrt{2\pi
A}}e^{iP.(r_e+r_h)/2} e^{i r_e\times r_h.\hat{z}/2}
e^{-(r_e-r_h-r_P)^2/4},
\end{equation}
\narrowtext
\noindent
where $\bbox{r}_e$ and $\bbox{r}_h$ are the electron and hole
positions, $\bbox{r}_P \equiv \hat{\bbox{z}}\times \bbox{P}
l^2/\hbar$, $A$ is the area of the system, and the vector potential
has been chosen in the symmetric gauge, $\bbox{A}(\bbox{r}) =
\bbox{B}\times\bbox{r}/2$.  We have chosen units in which $\hbar=l=1$,
and, in the following, we express energies in units of
$e^2/4\pi\epsilon\epsilon_0 l$. To make the discussion more
transparent, however, we reintroduce these units at appropriate
points.

The energies of the free exciton states, relative to the zero-point
kinetic energy of the electron and hole, are given by the expectation
values of the electron-hole interaction potential\cite{lernloz}. We
study a situation in which the electron and hole move in planes
separated by a distance $d$, such that the interaction is
\begin{equation}
\label{eq:veh}
V_d^{eh}(\bbox{r}) = - \frac{e^2}{4\pi\epsilon\epsilon_0}
\frac{1}{\sqrt{|\bbox{r}|^2+d^2}}.
\end{equation} 
The interaction energy of the exciton state $|\bbox{P}\rangle$ is
therefore
\begin{eqnarray}
\nonumber
E_d(\bbox{P}) & \equiv & \langle
\bbox{P}|V^{eh}_d(\bbox{r}_e-\bbox{r}_h)|\bbox{P}\rangle \\
 & = &
-\int_0^\infty e^{-u^2/2}e^{-ud}J_0(u|\bbox{P}|)\;du ,
\label{eq:generaldispersion}
\end{eqnarray}
where $J_0(z)$ is the ordinary Bessel function.  We have not found a
closed-form expression for this integral (in the limit $d=0$ it
reduces to the expression derived in Ref.\cite{lernloz}). However, in
what follows, we are interested in exciton states with wavevectors on
the scale of the reciprocal lattice vectors of the crystal, ${\cal
O}(1/a)$. At low filling fraction, this is much smaller than the scale
on which the energy~(\ref{eq:generaldispersion}) varies, ${\cal
O}(1/l)$, so for our purposes it is sufficient to work with an
expansion
\begin{equation}
\label{eq:quadraticdispersion} 
E_{d}(\bbox{P})\simeq -B_{d} + \frac{P^2}{2M_d} +{\cal O} (P^4),
\end{equation}
which may be interpreted in terms of a ``binding energy'', $B_{d}$,
and an ``effective mass'', $M_{d}$,
\begin{eqnarray}
B_d & = & \sqrt{\pi/2} \; e^{d^2/2}
\mbox{erfc}\left(d/\sqrt{2}\right)
\label{eq:generalbindingenergy}
\\
        & {\sim} & 
\sqrt{\pi/2} - d +1/2 \sqrt{\pi/2}d^2 + {\cal O} (d^3), \hspace{0.15cm} d\ll 1
\eqnum{5a}\label{eq:bindingenergysmall}\\
 & {\sim} &
    1/d + {\cal O} (1/d^2),  \hspace{0.15cm} d\gg 1
\eqnum{5b}\label{eq:bindingenergyasymptotic}\\
\label{eq:exactmass}
1/M_{d} & = & \sqrt{\pi/8} \; e^{d^2/2 }  (1 + d^2)
\;  \mbox{erfc}\left(d/\sqrt{2}\right)  - d/2 
\label{eq:generalmass}\\
          & {\sim} & 
 \sqrt{\pi/8} - d +3/2 \sqrt{\pi/8} d^2 + {\cal O} (d^3), \hspace{0.15cm} d\ll 1
\eqnum{6a}\label{eq:masssmall}\\
 & {\sim} &    1/d^3   + {\cal O} (1/d^4), \hspace{0.15cm} d\gg 1
\eqnum{6b}\label{eq:massasymptotic}
\end{eqnarray}

In many ways the exciton behaves as a simple free particle with a mass
$M_d$.  In particular, the velocity of an exciton in the state
$|\bbox{P}\rangle$ is $\partial
E(\bbox{P})/\partial\bbox{P}$\cite{gorkov}.  In view of this, it is
convenient to think of the contribution to $E_d(\bbox{P})$ which is
momentum-dependent as the ``kinetic energy'' of the exciton, even
though this energy originates from the electron-hole interaction.  The
exciton is overall charge neutral, but it does carry a dipole moment
and therefore couples weakly to an external electrostatic potential.
The dipole moment is of size $-ed\hat{\bbox{z}}$ perpendicular to the
plane of the interface, and of size $-e\bbox{r}_P$ parallel to this
plane. It is the dipole moment of the exciton that will cause it to be
scattered by the Wigner crystal of other electrons.  The strength of
the coupling of the exciton to a Wigner crystal with a lattice
constant $a$ is determined by the competition between the typical
dipole energy of the exciton, $d ^2 e^2/\epsilon a^3$, which is
minimized at an interstitial site of the lattice, and the kinetic
energy cost to confine the exciton to such a site, $1/(M_d a^2)$.
Using the asymptotic expression for the
mass~(\ref{eq:massasymptotic}), it is found that these two energies
become equal when $d/l\simeq 1/\nu^{1/10}$, where $\nu$ is the filling
fraction of the two-dimensional electron gas. This condition is so
weakly dependent on the filling fraction that it is accurate to say
that one expects the exciton states to crossover from being weakly
coupled to the lattice when $d\lesssim l$ to strongly confined to
interstitial regions when $d\gtrsim l$.

In order to understand the photoluminescence spectrum it is essential
to know both the energies of the exciton states and the rate at which
they decay to emit radiation.  Within the effective mass approximation
for the electron and hole bands, the radiative lifetime of an exciton
state depends on two factors: the band-to-band dipole matrix element
of the host semiconductor, and a matrix element between the electron
and hole envelope functions.  The first contribution is constant for
all transitions. We study only the envelope term, which is sufficient
to describe the relative recombination rates of different exciton
states.  The operator that describes electron-hole annihilation may be
written in second quantized notation as
\begin{equation}
\hat{L} = \sqrt{\frac{2\pi l^2}{A}}\int d^2\bbox{r}\;
\hat{\psi}_e(\bbox{r}) \hat{\psi}_{h} (\bbox{r}),
\end{equation}
where $\hat{\psi}_{h}$ and $\hat{\psi}_e$ are the field annihilation
operators for holes and electrons, and the wavevector of the emitted
photon has been assumed to be zero.  We have chosen a convenient
normalization, which is arbitrary since we consider only relative
radiative decay rates.  We define the ``luminescence strength'' of a
transition between any pair of (many-body) states $|i\rangle
\rightarrow |f\rangle$ by
\begin{equation}
\label{eq:lumstrengthdef}
L_{if} =  |\langle f| \hat{L} | i \rangle|^2.
\end{equation}
\widetext
\newpage
\narrowtext
\noindent
For the free exciton states discussed above~(\ref{eq:llwavefunctions})
the only available final state is the vacuum and we find
\begin{equation}
\label{eq:lumstrengthp}
L_P = |\langle \mbox{VAC}| \hat{L} | P\rangle|^2 = \delta_{P,0} .
\end{equation}
Thus, due to momentum conservation, only the $\bbox{P}=0$ exciton
state can emit a long-wavelength photon.  Since the free exciton
states form a complete set of states for the exciton in the lowest
Landau level, the luminescence strength of a transition in which an
exciton in a general state $|\psi\rangle$ decays to emit radiation may
be found from Eqn.~(\ref{eq:lumstrengthp}) to be given by
\begin{equation}
\label{eq:lumstrengthproject}
L_\psi = |\langle \bbox{P}=0|\psi\rangle|^2.
\end{equation}
We will use this expression in the following discussions which concern
the exciton states in an external potential, for which the momentum is
not a good quantum number.

The relative intensity of a photoluminescence transition is
proportional to its luminescence strength multiplied by the
probability for a photo-excited hole to be in the initial state of the
transition.  These probabilities are difficult to quantify: the
lifetime of the photo-excited hole can be shorter than its
equilibration time, so one may observe recombination from high-energy
non-equilibrium states.  Although we are primarily interested in
photoluminescence, we note here that the luminescence strength
$L_{if}$ also characterizes the strength of the transition, as
observed in optical absorption, from the state $|f\rangle$ to the
state $|i\rangle$. In this case, uncertainties related to
non-equilibrium populations do not arise as the lifetime of the hole
does not limit the time available for the initial states to
equilibrate.

\section{Exciton States in the Presence of a Wigner Crystal: 
Weak Coupling Limit}
\label{sec:weakcoupling}

We will now study the scattering of the exciton
states~(\ref{eq:llwavefunctions}) by the electrons forming the Wigner crystal.
As described in the introduction, we neglect the exchange interaction between
the electron in the exciton and those forming the crystal.  We anticipate that
this is a good approximation at low filling fraction, when the fraction of
basis states excluded from the exciton wavefunction by the Pauli exclusion
principle is small; this is confirmed by numerical studies in which the effects
of exchange are included\cite{cooperzhang}.  The model we discuss also
correctly represents a situation in which the spin or subband index of the
electron in the exciton is different from the corresponding index of the
electrons in the two-dimensional electron gas.

Neglecting exchange interactions, the exciton is scattered only by the
charge density
\begin{equation}
\label{eq:chargedensity}
\rho(\bbox{r}) = \sum_\alpha \left(\frac{-e}{2\pi l^2}
e^{-(\bbox{r}-\bbox{R}_\alpha)^2/2l^2} + \bar \rho\right)\;\delta (z),
\end{equation}
which represents a system of electrons in lowest Landau level orbitals
centered at the sites $\{\bbox{R}_\alpha\}$ of a triangular lattice
with a lattice constant $a$ (the filling fraction is therefore
$\nu=(4\pi/\sqrt{3}) l^2/a^2$).  The magnitude of $\bar{\rho}$ is
chosen to provide a uniform neutralizing positive background.  The
resulting external potential energies of the electron and hole are
\widetext
\top{-2.8cm}
\begin{equation}
\label{eq:staticpotential}
V^e(\bbox{r}_e)+V^h(\bbox{r}_h) = \int \rho(\bbox{r})
\frac{e}{4\pi\epsilon\epsilon_0} \left[
\frac{-1}{|\bbox{r}_e-\bbox{r}|} +
\frac{1}{\sqrt{(\bbox{r}_h-\bbox{r})^2+d^2}}\right] d^2\bbox{r}.
\end{equation}

The motion of the exciton in this potential is fully described by the
matrix elements of all interactions within the basis of free exciton
states~(\ref{eq:llwavefunctions})
\begin{equation}
H_{P^\prime,P} \equiv \langle \bbox{P}^\prime| V^{eh}(\bbox{r}_e-
\bbox{r}_h) + V^e(\bbox{r}_e) + V^h(\bbox{r}_h) |\bbox{P} \rangle.
\end{equation}
The first term is diagonal in this basis and gives rise to the free
exciton
energy~(\ref{eq:generaldispersion},\ref{eq:quadraticdispersion}).  The
last two terms describe the scattering. Due to the discrete
translational symmetry of the crystal, the only matrix elements which
are non-zero are those between states which differ by a reciprocal
lattice vector of the crystal, $\{\bbox{K}\}$.  Since we are
ultimately interested in states that can emit radiation, and which
therefore from Eqn.~(\ref{eq:lumstrengthproject}) must contain some
component of the zero-momentum free exciton state, we need only study
the basis states $|\bbox{K}\rangle$ in which the wavevector of the
exciton is a reciprocal lattice vector.  After a lengthy but
straightforward process we find
\begin{equation}
H_{K^\prime,K} = \delta_{K^\prime,K}\; E_{d}(\bbox{K}) + (1-
\delta_{K^\prime,K})\nu \frac{ e^{-3(K-K^\prime)^2/4}}{|K-K^\prime
|} \left[ e^{i\bbox{K}\times \bbox{K}^\prime.\hat{\bbox{z}}/2} -
e^{-|K-K^\prime |d} e^{-i\bbox{K}\times
\bbox{K}^\prime.\hat{\bbox{z}}/2} \right].
\label{eq:matrixelements}
\end{equation}

While the above expressions for the matrix elements are exact and
therefore fully describe the behaviour of our model, in their present
form they are not useful for analytic purposes.  To obtain some
insight into the properties of this system we therefore introduce some
approximations.  In what follows, we develop a perturbation theory in
$d/l$, based on the free exciton states, which is valid in the limit
of low filling fraction.

\newpage
\narrowtext

Consider the case $d=0$ in which the hole moves in the same plane as
the electrons.  In this limit, the zero-momentum exciton state is not
scattered by the potential, and is therefore an exact energy eigenstate
\begin{equation}
H_{K^\prime,K=0} = E_{d=0}(0)\;\delta_{K^\prime, 0} .
\end{equation}
All other energy eigenstates must be orthogonal to this state, so from
Eqn.~(\ref{eq:lumstrengthproject}) they must have zero luminescence
strength.  There is therefore only a single line in the
photoluminescence spectrum. Moreover, this line appears at the same
energy as the exciton line in the absence of the two-dimensional
electron gas.  This recovers a restricted version of the general
result that when electrons and holes are confined to the lowest Landau
level and move in the same plane, photoluminescence contains no
spectroscopic information on the state of the two-dimensional electron
gas\cite{macrezkell,apalkovprb}.

Although the free exciton state with zero momentum is an exact energy
eigenstate of the system at $d=0$, other free exciton states are
not. There do remain non-zero off-diagonal matrix elements coupling
these states.  However, in the limit of small filling fraction, this
coupling may be neglected for the lowest energy states: states with
small momenta $|\bbox{K}|\sim 1/a$, are only coupled by off-diagonal
terms of order $(l/a)^3$, whereas the energy spacing between these
states is of order $\hbar^2 /(M_{d=0}a^2)\sim (l/a)^2$.  We choose to
work in an expansion in small filling fraction, and neglect terms of
order $(l/a)^3$ and higher.  Within this approximation, the free
exciton states are eigenstates of the Hamiltonian at $d=0$.

We now study the deviations from $d=0$ within a perturbation expansion
in the separation $d$.  Expanding the Hamiltonian in this parameter,
we find
\widetext
\top{-2.8cm}
\begin{equation}
H_{K^\prime,K} \simeq  \delta_{K^\prime, K} \left[ - B_d +
\frac{K^2}{2} \left( \sqrt{\frac{\pi}{8}}-d
+\frac{3}{2}\sqrt{\frac{\pi}{8}}d^2 \right) \right] 
 + (1 -
\delta_{K^\prime,K}) \nu d +{\cal O}(1/a^3) +{\cal O}(d^3/a^2).
\label{eq:matrixelementsnud}
\end{equation}
\bottom{-2.7cm}
\narrowtext
\noindent
To lowest order in $d$ all states are coupled by a matrix element~$(1
-\delta_{K^\prime, K})\nu d$. This leads to a mixing of the radiative
state, $|\bbox{K}=0\rangle$, with all other exciton states of
reciprocal lattice vector; if these states become populated on
photo-excitation, each will contribute an additional line to the
photoluminescence spectrum.  The strongest new line arises from the
six lowest-lying energy states, at $|\bbox{K}|=4\pi/(\sqrt{3}a)$.
Within degenerate perturbation theory, these states are split into a
five-fold degenerate level shifted in energy by $-\nu d$, and a single
state shifted by $+5\nu d$.  We now apply first-order perturbation
theory to calculate the mixing of $|\bbox{K}=0\rangle$ into these
states. This amplitude is zero for the five degenerate states, and
equal to $\sqrt{6}\nu d$ for the single state split off from these
five.  The luminescence spectrum therefore consists of a strong
spectral line from what was the free zero-momentum exciton state, at
an energy (relative to the band-gap plus zero point kinetic energy)
and with a luminescence strength
\begin{eqnarray}
E_0 & = & - B_d, \\ L_0 & = & 1 - {\cal
O}(d^2)
\end{eqnarray}
and an additional line arising from the exciton states with the
smallest non-zero reciprocal lattice vector, with an energy and
luminescence strength
\begin{eqnarray}
E_1 & = & - B_d + \frac{(4\pi/\sqrt{3}a)^2}{2}\left[
\sqrt{\frac{\pi}{8}}-d + \frac{3}{2}\sqrt{\frac{\pi}{8}} d^2 \right] +
5\nu d \\ L_1 & = & \left|\frac{\sqrt{6}\nu
d}{\frac{1}{2}\sqrt{\frac{\pi}{8}} (4\pi/\sqrt{3}a)^2} \right|
^2 = \frac{36}{\pi^3}d^2  .
\end{eqnarray}
The next line to higher energy arises from the six states at
$|\bbox{K}|=4\pi/a$, and has a luminescence strength that is smaller
than this by a factor of $[(4\pi/\sqrt{3})/(4\pi)]^4=1/9$.

Thus, for small $d$ the presence of the crystal causes peaks to appear
in the photoluminescence spectrum to higher energy than the main
exciton line.  The energy spacing between these peaks is primarily
determined by the effective mass of the exciton and the reciprocal
lattice vectors of the Wigner crystal, with a small correction due to
the coupling to the lattice potential.  The luminescence strengths of
these higher energy exciton states are small compared to the lowest
energy exciton transition, and will therefore give rise to much weaker
spectral features.  The luminescence strengths of these features grow
as $d$ increases.  However, at the same time, the exciton becomes more
strongly coupled to the lattice potential, so the energies of these
states deviate from the energies of the reciprocal lattice vector
excitons.  Once the coupling is larger than the spacing between the
free exciton states, $\nu d\gtrsim 1/M_da^2$, the corrections to these
energies become large, and the above perturbation theory breaks down.
This condition is equivalent to the strong-coupling condition,
$d\gtrsim l$, that we derived earlier. In the strong-coupling limit, a
quite different approach is required to describe the low energy
exciton states.

\section{Exciton States in the Strong Coupling Limit: The Interstitial Exciton} 
\label{sec:interstitial}

We now turn to the limit of strong exciton-lattice coupling, in which
the kinetic energy of the exciton is small compared to its dipole
potential energy in the lattice.  If we were to neglect the kinetic
energy completely, then the groundstate of the system would be the
same as that of the classical groundstate of electrons in the presence
of an ionized donor impurity\cite{classicalruzin}: the exciton, with
its weak dipole moment, would position itself at an interstitial site
of the lattice, and, for $d<0.29 a$, the remaining electrons would
form an essentially undeformed Wigner crystal.  The exciton does,
however, have some residual dynamics. In this section we develop an
effective Hamiltonian that describes the motion of the centre of the
exciton in a smooth external potential.  We then apply this to the
potential close to an interstitial site of the Wigner
crystal~(\ref{eq:staticpotential}), and study the energies and
luminescence strengths of the low energy states.  Those who do not
care to follow the derivation of the effective theory may proceed to
section~\ref{subsec:harmonic} without loss of continuity.
\widetext
\top{-2.8cm}
\subsection{Effective Hamiltonian} \label{sub:effectivehamiltonian}

In order to derive an effective Hamiltonian for the motion of an
exciton in a smooth external potential, we extend the approach of
Gor'kov and Dzyaloshinskii\cite{gorkov} to include the potential
energies $V^e(\bbox{r}_e)$ and $V^h(\bbox{r}_h)$ felt by the electron
and hole.  We start from the full Hamiltonian for the interacting
electron-hole pair in a uniform magnetic field
\begin{equation} 
\label{eq:fullhamiltonian} 
H = \frac{(
\bbox{p}_e + e\bbox{A}_e)^2}{2m_e}
 + \frac{( \bbox{p}_h - e\bbox{A}_h)^2}{2m_h} + V^{eh}
 (\bbox{r}_e - \bbox{r}_h) +V^e(\bbox{r}_e)
 +V^h(\bbox{r}_h), 
\end{equation} 
and work in the symmetric gauge.  We transform to a new set of
co-ordinates, the first of which is the momentum defined in
Ref.\cite{gorkov}; the remaining co-ordinates are the position of the
centre of mass and a momentum and position describing the internal
motion
\begin{eqnarray}
\label{eq:ganddmomentum}
\bbox{P} & \equiv & \bbox{p}_e + \bbox{p}_h - \frac{e}{2}
\bbox{B}\times (\bbox{r}_e - \bbox{r}_h) \\
\label{eq:cofm}
\bbox{R}_c & \equiv & \frac{m_e\bbox{r}_e + m_h \bbox{r}_h}{m_e+m_h} \\
\label{eq:internalmomentum}
\bbox{p} & \equiv &  \frac{m_h\bbox{p}_e -
m_e\bbox{p}_h}{m_e+m_h} +
\frac{e}{2} \bbox{B}\times \left( \frac{m_e\bbox{r}_e + m_h
\bbox{r}_h}{m_e+m_h} \right)\\
\bbox{r} & \equiv & \bbox{r}_e - \bbox{r}_h .
\end{eqnarray}
The centre-of-mass and the internal co-ordinates behave as independent,
canonically conjugate pairs.  In terms of these new co-ordinates, the
Hamiltonian is
\begin{equation}
H = \frac{( \bbox{P} + e\bbox{B}\times \bbox{r}
)^2}{2 M} + \frac{( \bbox{p} + \gamma e/2 \;\bbox{B}\times
 \bbox{r} )^2}{2 \mu} + V^{eh} (\bbox{r}) + V^{e}
 (\bbox{R}_c+ \eta_h \bbox{r}) + V^{h} (\bbox{R}_c-
 \eta_e \bbox{r}),
\end{equation}
\bottom{-2.7cm}
\narrowtext
\noindent
where $M\equiv m_e+m_h$, $\mu\equiv m_em_h/(m_e+m_h)$, $\gamma \equiv
(m_h-m_e)/(m_e+m_h)$, and $\eta_\alpha \equiv m_\alpha/(m_e+m_h)$.

First consider the free exciton, $V^e=V^h=0$. The Hamiltonian is
independent of $\bbox{R}_c$, so the momentum, $\bbox{P}$, is conserved
and can be replaced by its eigenvalue to leave a Hamiltonian for the
internal co-ordinates, $\bbox{r}$ and $\bbox{p}$.  If the interaction
potential, $V^{eh}(\bbox{r})$, is neglected in relation to the kinetic
energy, the internal motion consists of Landau level orbits centred on
$\bbox{r}_P\equiv\hat{\bbox{z}}\times\bbox{P} l^2/\hbar$.  This is the
neglect of Landau level coupling, and is the simplification used in
Refs.\cite{lernloz,gorkov}; in the lowest Landau level, it leads to
the states that we have discussed above~(\ref{eq:llwavefunctions}).

We now introduce the potentials, $V^e$ and $V^h$, and follow a similar
approximation. The momentum is no longer conserved. However, provided
the potential is sufficiently weak, the momentum only changes slowly
compared to the rapid cyclotron motion. The motion can then be treated
adiabatically, with the fast internal motion fixed in the lowest
Landau level and adjusting to follow the slowly changing momentum.
This procedure is analogous to the Born-Oppenheimer approximation in
the theory of molecular dynamics\cite{llqm}, where the fast
electronic degrees of freedom are eliminated to provide an effective
theory for the slow atomic co-ordinates. In the same way, we obtain an
{\it effective Hamiltonian} for the operators $\bbox{P}$ and
$\bbox{R}_c$.  Before we write down this Hamiltonian, we make one last
change of variables. It is convenient to work is terms of the
co-ordinate $\bbox{R} \equiv (\bbox{r}_e +\bbox{r}_h)/2$ rather than
the centre of mass. This avoids any irrelevant mass dependence in our
analysis (``irrelevant'' since both electron and hole are restricted
to the lowest Landau level).  This transformation does not affect the
commutation relation of the position and momentum co-ordinates.  The
effective Hamiltonian in terms of these co-ordinates is
\widetext
\begin{equation}
H = \hbar \omega_\mu/2 + E_d(\bbox{P}) + \int \left[ V^e(\bbox{R} +
(\bbox{r}^\prime + \bbox{r}_{P})/2) + V^h(\bbox{R} - (\bbox{r}^\prime
+ \bbox{r}_{P})/2) \right] \frac{e^{-{r^\prime}^ 2/2l^2}}{2\pi l^2} \;
d^2\bbox{r}^\prime,
\label{eq:integral}
\end{equation}
where $\hbar \omega_\mu/2=\hbar eB/2\mu$ is the zero-point kinetic
energy of the electron and hole, and $E_d(\bbox{P})$ is the
dispersion relation~(\ref{eq:generaldispersion}) arising from their
mutual attraction.

To convert this expression into a more convenient form, we make use of
the simplifications available in the problem.  As we are interested in
small momenta, ${\cal O}( 1/a) \ll 1/l$, we expand the dispersion
relation to quadratic order. Also, provided the exciton is not very
close to a lattice site, the potentials $V^e$ and $V^h$ are smooth on
the lengthscale of the magnetic length, and it is a good approximation
to expand these terms in the integral of Eqn.(\ref{eq:integral}) to
first order in $(\bbox{r}^\prime + \bbox{r}_P)$. We find
\begin{equation} 
\label{eq:finalform}  
H \simeq \hbar \omega_\mu/2 - B_d + \frac{\bbox{P}^2}{2 M_d} + V^{e}
(\bbox{R}) + V^{h} (\bbox{R}) - \frac{1}{2} \bbox{P}.
\left[\bbox{\nabla} V^{e} (\bbox{R}) - \bbox{\nabla} V^{h}(\bbox{R})
\right]\times \bbox{\hat{z}}.
\end{equation}
This is a general expression for the motion of the exciton in smooth
external potentials, and in a strong magnetic field.  An expression
for the mass $M_d$ was calculated earlier~(\ref{eq:exactmass}).  Since
the position and momentum operators are canonically conjugate
co-ordinates, the energy eigenstates of the exciton follow from the
solutions of the Schr\"{o}dinger equation with the Hamiltonian
(\ref{eq:finalform}); it is to be understood that the final term of
this expression is symmetrized in the position and momentum operators,
such that the Hamiltonian is Hermitian. Two approximations were used
to derive this expression. Firstly, the rapid cyclotron motion was
assumed to adiabatically follow the changing momentum; this is valid
provided the spacing between the energy levels arising from the
centre-of-mass motion is small compared to the cyclotron energy,
$\hbar\omega_\mu$ (no Landau level coupling). Secondly, we expanded
the external potential to first order in $(\bbox{r}^\prime +
\bbox{r}_P)$, neglecting terms of order $l^2\nabla^2 V$.

\subsection{Motion in the Wigner Crystal: Harmonic Approximation}
\label{subsec:harmonic}

We now use the formalism developed in the previous subsection to study
the motion of the exciton in a Wigner crystal.  We therefore introduce
the electrostatic interactions~(\ref{eq:staticpotential}) as the
external potentials in the effective Hamiltonian~(\ref{eq:finalform}).
This expression for the effective Hamiltonian requires the external
potential to vary slowly on the scale of the exciton size, $l$. We
therefore expect this formalism to provide an appropriate description
of the low-energy states when $d\gtrsim l$, in which the exciton
remains far from the interstitial sites. However, we still assume that
$d/a\ll 1$, and expand the effective Hamiltonian in this parameter to
find 
\begin{equation} 
H = \frac{\bbox{P}^2}{2M_d} +
\frac{e^2}{8\pi\epsilon\epsilon_0} \sum_\alpha
\frac{d^2}{|\bbox{R}-\bbox{R}_\alpha|^3}  +
\frac{e^2}{4\pi\epsilon\epsilon_0}\frac{l^2
\bbox{P}}{\hbar}.\sum_\alpha\hat{\bbox{z}}\times\frac{\bbox{R}-\bbox{R}_\alpha}{|\bbox{R}-\bbox{R}_\alpha|^3},
\label{eq:adiabatic} 
\end{equation}
\bottom{-2.7cm}
\narrowtext
\noindent
where the sums run over the triangular lattice sites
$\{\bbox{R}_\alpha\}$, and the final term should be interpreted as the
symmetrized product of the momentum-dependent and position-dependent
factors. We have dropped the constant binding energy, $-B_d$, and the
zero-point kinetic energy, $\hbar\omega_\mu/2$, and have also
neglected a constant energy, $-\nu d$, which arises from the dipole
moment of the exciton in the electric field of the neutralizing
positive background.  Each of the terms appearing in this effective
Hamiltonian has a simple intuitive interpretation.  The first two
terms simply represent the kinetic energy of the exciton and the
dipole energy for the exciton to be centred at the position
$\bbox{R}$.  The final term is less familiar, but also arises from a
dipole energy: in this case, due to the in-plane dipole moment of the
exciton, which is of a size $-e\bbox{r}_P$ for an exciton with
momentum $\bbox{P}$. The resulting contribution to the Hamiltonian
resembles the first-order coupling of a charged particle to a vector
potential.

We have argued that for large $d/l$ the exciton is confined at the
interstitial sites of the triangular lattice.  We now study the
low-lying states of this effective Hamiltonian within the simplest
approximation, in which the potential is expanded to harmonic order
about an interstitial site.  Keeping only the contributions to the
confinement potential arising from the nearest three lattice
electrons, we find
\begin{equation}
H = \frac{\bbox{P}^2}{2M_d} +
\frac{1}{2}\frac{243\sqrt{3}}{4}\frac{d^2}{a^5}\bbox{R}^2
-\frac{9\sqrt{3}}{2} \frac{1}{a^3}
\left(\bbox{R}\times\bbox{P}\right).\hat{\bbox{z}},
\end{equation}
plus a constant energy of $11.57d^2/a^3$ due to the dipole energy at
the interstitial site of the triangular
lattice\cite{classicalruzin}. We have again set
$e^2/4\pi\epsilon\epsilon_0 l = \hbar=l=1$. Due to the rotational
invariance of this Hamiltonian, its eigenstates may be classified by
their angular momenta, $\bbox{R}\times\bbox{P}.\hat{\bbox{z}}$.  We
restrict attention to states with {\it zero} angular momentum, since
only these states can emit long-wavelength radiation.  The zero
angular momentum states of the two-dimensional harmonic oscillator
have the wavefunctions and energies
\begin{eqnarray}
\label{eq:showavefunctions}
\psi^{sho}_n(\bbox{R}) & = & \frac{1}{\sqrt{\pi
R_0^2}}e^{-R^2/2R_0^2}{\cal L}_n(R^2/R_0^2),\\
\label{eq:shoenergy}
E_n & = & \hbar\Omega (2n+1),
\end{eqnarray}
where
\begin{eqnarray}
\label{eq:shoquantum}
\hbar\Omega & = & \sqrt{\frac{243\sqrt{3}}{4}
\frac{d^2}{a^5} \frac{e^2}{4\pi\epsilon\epsilon_0}\frac{\hbar^2}{M_d}}
\\
& \stackrel{d\gg l}{\longrightarrow} & \sqrt{\frac{243\sqrt{3}}{4}
\frac{l^6}{a^5 d}} \frac{e^2}{4\pi\epsilon\epsilon_0l}, 
\eqnum{33a}\label{eq:shoquantumasym}\\
\label{eq:sholength}
R_0 & = & \left(\frac{4}{243\sqrt{3}}\frac{4\pi\epsilon\epsilon_0}{e^2}
\frac{\hbar^2 a^5}{d^2 M_d}\right)^{1/4} \\
 & \stackrel{d\gg l}{\longrightarrow} & \left(\frac{4}{243\sqrt{3}}
\frac{a^5}{d^5}\right)^{1/4} l,
\eqnum{34a}\label{eq:sholengthasym}
\end{eqnarray}
and ${\cal L}_n(z)$ are the Laguerre polynomials.  The expressions for
$d\gg l$ follow from the asymptotic expansion of the effective
mass~(\ref{eq:massasymptotic}).  Note that the two-dimensional
harmonic oscillator states with zero angular momentum are spaced in
energy by {\it twice} the oscillator quantum, $\hbar\Omega$.

Radiative recombination occurs from each of these states.  For
consistency with the definition of the luminescence strength in the
weak coupling limit, we construct states consisting of a superposition
of harmonic oscillator states, placed at each interstitial site of a
crystal with an area $A$. The only such linear combination with
non-zero luminescence strength is the state in which all states are
combined with the same amplitude and phase.  From
Eqn.(\ref{eq:lumstrengthp}) we find that the luminescence strength of
such a state constructed from $\psi^{sho}_n$ is
\begin{equation}
\label{eq:sholuminescence}
L_n \equiv \frac{4}{\sqrt{3}a^2}\left|\int d^2\!\bbox{R}\;
\psi^{sho}_n(\bbox{R})\right|^2 = \frac{16\pi R_0^2}{\sqrt{3} a^2}.
\end{equation}
Note that this is independent of the index $n$, so the ground state
and all excited states have the same luminescence
strengths\cite{footnote}. Such behaviour contrasts with the
weak-coupling limit for which the lowest energy exciton state is much
more strongly coupled to radiation than the higher energy states. As
the spacing $d$ is increased from zero, so that the system evolves
from the weak-coupling to the strong-coupling limit, the higher energy
exciton transitions grow in luminescence strength from zero (optically
inactive) to eventually attain the same strength as the lowest-energy
exciton state.

In summary, within the harmonic approximation for the potential at an
interstitial site of the lattice, the exciton contribution to the
photoluminescence spectrum consists of a series of lines which are
uniformly spaced and have relative intensities given by the relative
populations of the states.  This approximation is valid provided the
typical size of the exciton state is small compared to the lattice
constant.  From equation~(\ref{eq:sholength}), one finds that this
depends only very weakly on the filling fraction, and occurs when
$d\gtrsim l$.  This is the same as the strong-coupling condition under
which the perturbation expansion of the previous section failed.

\section{Experimental Comparisons}
\label{sec:comparison}

Several groups have reported studies of the itinerant-hole
photoluminescence spectra of the two-dimensional electron systems
formed in high-mobility GaAs/AlGaAs
devices\cite{clarkreview,goldys,heiman,turberprl,turberfieldeuro}. Measurements
on single heterojunctions and single quantum wells show qualitatively
the same behaviour in photoluminescence: in the fractional quantum
Hall regime two spectral lines are observed; as the filling fraction
is reduced below about $\nu=1/6$, the higher energy peak is found to
split to form a doublet.  The filling fraction below which this
doublet structure appears correlates well with the filling fraction at
which transport measurements show a transition to an insulating
state\cite{goldys}. This has motivated claims that the doublet is
associated with the formation of a magnetically induced Wigner
crystal\cite{goldys,turberfieldeuro}.  However, the origin of this
doublet is still not well understood.  In this section, we discuss
whether this doublet can be accounted for in terms of the structure
that we predict.  We will compare the energy splitting of the doublet
with the energy difference between the two lowest-lying exciton states
of our model that are optically active.

For numerical comparisons of our theory with experiment, we will focus
on the results of Goldys~{\it et~al.} presented in Ref.\cite{goldys}.
This paper reports studies of the photoluminescence spectrum of a high
mobility GaAs/AlGaAs heterostructure with density $3.2\times
10^{10}\mbox{cm}^{-2}$.  Specifically we will concentrate on a field
of $B=12.5\mbox{T}$ which is appropriate for Fig.~4(a) of
Ref.\cite{goldys} for which the doublet structure, with a splitting of
0.5meV, is well developed.  Under these conditions, the filling
fraction is $\nu=0.1$, corresponding to a magnetic length of
$l=72\mbox{\AA}$ and a lattice constant of $a=600\mbox{\AA}$ for the
triangular Wigner crystal.

The most uncertain parameter that enters our model is the spacing $d$.
This distance is expected to depend on the details of the band-bending
in the vicinity of the interface, and may even vary with magnetic
field\cite{vietbirman}.  We will assume values of $d=50\mbox{\AA}$ and
$d=100\mbox{\AA}$ as small and large estimates for this quantity.
These are consistent with recent numerical studies of the binding of
an exciton to the interface of a
single-heterojunction\cite{vietbirman}.  Note that these values are
also consistent with the condition $d< 0.29a$ which is required by our
theory in order that the exciton does not strongly deform the crystal.

The smaller of these two separations, $d=50\mbox{\AA}$, is less than
the magnetic length, so we expect our weak-coupling formulae to be
appropriate.  In the weak coupling limit, the difference in energy
between the main exciton line and the lowest-energy optically-active
state is primarily determined by the free exciton energy at a
wavevector equal to the smallest non-zero reciprocal lattice vector
$|\bbox{K}|=4\pi/(\sqrt{3}a)$.  Taking the dielectric constant of GaAs
to be $\epsilon=12.53$, we find from
equation~(\ref{eq:generaldispersion}) that the energy spacing is
1.3meV.

The larger spacing, $d=100\mbox{\AA}$, satisfies $d>l$, and we should
therefore use our strong-coupling formulae, which state that the
energy spacing between adjacent optically-active exciton states is
$2\hbar\Omega$ where $\hbar\Omega$ is given by
equation~(\ref{eq:shoquantum}).  Using the appropriate value for the
effective mass of the exciton (\ref{eq:exactmass}), this evaluates to
an energy spacing of 0.7meV at $d=100\mbox{\AA}$. Under these
conditions, the characteristic size of the groundstate wavefunction is
$R_0=150\mbox{\AA}$, which is smaller than, but comparable to, the
distance to the closest saddle point of the crystal potential,
$a/\sqrt{12} = 170\mbox{\AA}$.  This indicates that, under these
conditions, the harmonic approximation is at the limit of its
validity.

An exact calculation of the energy spacings predicted by our model for
these two cases is likely to lead to energies that are slightly
smaller than the above estimates\cite{cooperzhang}: deviations from
the weak coupling limit will tend to push the energy spacing towards
the (smaller) energy spacing one would obtain by using the
strong-coupling theory; away from the strong-coupling limit, the use
of the harmonic approximation will over-estimate the strength of the
confinement potential and hence also overestimate the energy spacing.
We also expect a small reduction of both energies due to the finite
thicknesses of the electron and hole subband wavefunctions in a real
device.  Even allowing for these reductions, the resulting energy
spacings remain larger than the expected zone-boundary magnetophonon
energy, $0.2\mbox{meV}$\cite{cotemacdonald}. Our neglect of the
dynamical properties of the Wigner crystal is therefore a consistent
assumption.

The energy separation between the lowest two optically active states
in our model is on the same scale as the observed energy splitting of
the doublet, 0.5meV\cite{goldys}. It is therefore possible that this
structure is due to the splitting of the exciton peak as described by
our model.  Note that, if it is indeed correct to attribute this
structure to the predictions of our model, one must assume that the
experimental system is in the strong-coupling limit, in which the
exciton is strongly confined to an interstitial site.  There are two
important reasons for this, in addition to the fact that the
experimental energy spacing (0.5meV) compares more favourably with a
spacing of $d=100\mbox{\AA}$ than with smaller values.  Firstly, it is
observed that the intensities and the radiative lifetimes of the two
peaks forming the doublet are similar\cite{goldys}. The luminescence
strengths of the two transitions must therefore be comparable. In the
weak coupling limit, all higher energy exciton states have a much
smaller luminescence strength than the lowest energy exciton line; one
must go to the strong-coupling limit before these become equal [see
equation~(\ref{eq:sholuminescence}) and the following discussion].
Secondly, the observed energies of the two transitions forming the
doublet are independent of the temperature of the substrate over the
range $0.1 - 3$K\cite{goldys}. Since the Wigner crystal is expected to
melt at a temperature of 0.3K at these densities\cite{morf}, one
cannot attribute this structure to a nearly-free exciton state: the
energy of this state would be sensitive to the changing long-range
correlations at the melting transition.  Instead, it is appropriate to
attribute this structure to a strongly-confined interstitial exciton,
which is only sensitive to the short-range correlations of the
electron gas.

Although it is possible that the doublet structure is due to the
splitting of the exciton transition, as described by our model, there
are certain problems with this interpretation.  Firstly, it is not
clear that one should even observe any measurable signal from the
higher energy exciton states: these states lie more than 0.5meV above
the lowest energy exciton transition, so, if the exciton were to be in
thermal equilibrium, the expected populations of these states would be
vanishingly small. Secondly, assuming that there is some
non-equilibrium population of the higher energy exciton states, one
would perhaps expect to see more than one additional line (recall that
for a strongly-confined interstitial exciton all excited states have
the same luminescence strength).  Furthermore, this is not the only
possible explanation for this structure.  For instance, the
recombination from a negatively charged exciton, which has been
resolved in low-density GaAs wells at weak
field\cite{finkelstein,shieldsquenching} and in CdTe quantum wells in
strong magnetic field\cite{khengprl}, may also be expected to appear
in these systems, and has not as yet been identified.  We are
therefore cautious to ascribe the observed doublet to the line
splitting predicted by our model, which may not appear in these
experiments due to a lack of population of the higher energy exciton
states.  Further experiments are required in order to identify this
structure.  In particular, a study of the optical absorption spectrum
would be very valuable in this respect.  The exciton transitions that
we have discussed will present much stronger spectral features in this
experiment than in photoluminescence since the population factors,
which reduce the observed intensities of high-energy exciton states in
photoluminescence, are removed.

\widetext
\newpage
\narrowtext

\section{Summary}
\label{sec:conclusions}

We have presented a model of photoluminescence in the presence of a
magnetically-induced Wigner crystal, for systems in which the
photo-excited hole lies close to the interface.  This model predicts a
splitting of the exciton transition due to the scattering of the
exciton by the electrons forming the Wigner crystal.  We studied the
behaviour of our model as a function of the separation $d$.  For $d<
l$, the exciton is spread over many unit cells of the crystal.
Additional peaks with small luminescence strength can appear in the
photoluminescence spectrum, arising from higher energy states in which
the exciton has a wavevector equal to a reciprocal lattice vector of
the crystal.  The spectrum carries diffraction information and is
therefore sensitive to long-range crystalline order.  For $d>l$, the
exciton becomes strongly confined to an interstitial region.  We
studied this limit by developing an effective theory for the
centre-of-mass motion of the exciton.  Within the harmonic
approximation for the potential at an interstitial site, the spectrum
consists of an equally spaced set of peaks with relative intensities
determined by the relative populations of the various states.

We discussed the photoluminescence experiments of Ref.\cite{goldys}
and showed that the energy splitting of the doublet which has been
associated with the presence of a Wigner crystal is comparable to the
energy splitting we expect from our model for $d\simeq 100\mbox{\AA}$.
We argued that this could be due to the recombination from a
interstitial exciton.  However, other explanations are also possible,
and further experimental investigation is required to identify the
structure our model predicts.  In particular, this could best be
observed in optical absorption spectroscopy.
\vskip 0.3cm

Whilst preparing this work for publication, we learned of work by
J.~R.~Chapman, N.~F.~Johnson and V.~N.~Nicopoulos, proposing an
interpretation of the photoluminescence experiments that is somewhat
different from that presented here\cite{chapmanpreprint}; we are
grateful to them for a preprint of their work.

\acknowledgements{I am grateful to R.G.~Clark, A.~Turberfield, and
D.~Heiman for stimulating my interest in this problem and for helpful
subsequent discussions.  I would also like to thank Dr John Chalker
for much help with the development and presentation of this work, and
Prof. Bertrand Halperin and the Physics Department at Harvard
University for their hospitality and for many useful discussions on
this and related topics.  This work was supported in part by DENI and
in part by the NATO Science Fellowship Programme. }

\widetext


\begin{thebibliography}{10}
\bibitem{footnote} This can only apply for harmonic oscillator states
with $n$ less than $n_{\mbox{\tiny max}} \sim a^2/R_0^2$, which have a
size small compared to the lattice constant.  For higher energy states
the harmonic approximation fails: the luminescence strengths of these
states will be less than that predicted by the harmonic approximation.
\end{thebibliography}
\end{document}